\documentclass[iop]{emulateapj}

\newcommand{\pref}{\protect\ref}

\newcommand{\hinode}{{\em Hinode{}}}
\newcommand{\iris}{{\em IRIS{}}}
\newcommand{\sdo}{{\em SDO{}}}
\newcommand{\stereo}{{\em STEREO{}}}
\newcommand{\degree}{${^\circ{}}$}

\shorttitle{Solar Cycle 25}
\shortauthors{McIntosh \& Leamon}

\begin{document}

\title{Deciphering Solar Magnetic Activity: Spotting Solar Cycle 25}

\author{Scott W. McIntosh}
\affil{High Altitude Observatory,\\
National Center for Atmospheric Research,\\
P.O. Box 3000,\\
Boulder, CO 80307, USA}

\author{Robert J. Leamon}
\affil{Department of Astronomy,\\
University of Maryland,\\
College Park, MD 20742, USA}

\begin{abstract}
We present observational signatures of solar cycle 25 onset. Those signatures are visibly following a migratory path from high to low latitudes. They had starting points that are asymmetrically offset in each hemisphere at times that are 21-22 years after the corresponding, same polarity, activity bands of solar cycle 23 started their migration. Those bands define the so-called ``extended solar cycle.'' The four magnetic bands currently present in the system are approaching a mutually cancelling configuration, and solar minimum conditions are imminent. Further, using a tuned analysis of the daily band latitude-time diagnostics, we are able to utilize the longitudinal wave number ($m=1$) variation in the data to more clearly reveal the presence of the solar cycle 25 bands. This clarification illustrates that prevalently active longitudes (different in each hemisphere) exist at mid-latitudes presently, lasting many solar rotations, that can be used for detailed study over the next several years with instruments like the Spectrograph on \iris{}, the Spectropolarimeter on {\em Hinode}, and, when they come online, similar instruments on the Daniel K. Inouye Solar Telescope (DKIST) as we watch those bands evolve following the cancellation of the solar cycle 24 activity bands at the equator late in 2019.
\end{abstract}


\section{Introduction} \label{intro}
%

The waxing and waning number of the Sun's eponymous spots on decadal timescales has been a source of vigorous investigation since the discovery of the ``solar cycle'' in 1844 \citep[e.g.,][]{1844AN.....21..233S}. Sixty years later it was noted that, as the number of spots swells to solar maximum and shrinks to solar minimum over the course of 11(\--ish) years, their latitudinal distribution follows a migratory path from mid-latitudes (about $\pm$35\degr) to their eventual disappearance near the equator \citep[][]{1904MNRAS..64..747M}. Following a couple of years of relative calm at solar minimum the spots appear again at mid-latitudes with great abandon and the progression to the equator starts afresh \-- defining the start of the next solar cycle and giving the progressing latitudinal spot distribution the appearance of butterfly wings. Since the start of the last century the explanation (and prediction) of the 11(\-ish) year variability {\em with} the evolution of the so-called ``butterfly diagram'' has become a fundamental issue in solar and stellar physics. 

Over the last few years a new observational diagnostic technique has been applied to the understanding of solar variability \citep[][]{2014ApJ...792...12M}. Ubiquitous small features observed in the Sun's extreme-ultraviolet corona, ``EUV Brightpoints,''  \citep[or BPs; e.g.,][]{1974ApJ...189L..93G, 2003ApJ...589.1062H, 2005SoPh..228..285M} have been associated with tracing the evolution of the rotationally-driven giant convective scale \citep[][]{2014ApJ...784L..32M} that had vertices that were dubbed ``g-nodes.'' Together, these features permit the tracking of the magnetic activity bands of the 22-year magnetic cycle of the Sun that extend the conventional picture of decadal-scale solar variability, or the ``extended solar cycle'' as it has become known \citep[e.g.,][]{1988Natur.333..748W}. \citet{2014ApJ...792...12M} inferred that the global-scale (intra- and extra-hemispheric) ``telecommunication'' of these long-lived magnetic bands plays the critical role in the production of sunspots on those bands and thus define the (temporal) landmarks of the solar cycle.

\cite{2014ApJ...792...12M} noted that new cycle sunspot onset followed a time when the low-latitude pair of bands abruptly ``terminate'' at the equator, for example, the cycle 23 sunspots did not appear to grow in abundance or size until the cycle 22 bands had terminated (in mid-1997). Similarly, the polarity mirror-image of this progression occurred in early 2011 for cycle 24 sunspots,  following the termination of the cycle 23 bands. This equatorial termination, or cancellation, appears to signal the end of one sunspot cycle and leaves only the higher-latitude band in each hemisphere. Sunspots rapidly appear and grow on that mid-latitude band for several years in this, the ``ascending phase,'' until the next (oppositely-signed) band appears at high latitude. The presence of the new oppositely signed band triggers a downturn in sunspot production on the mid-latitude band; this occurrence defines the maximum activity level of that band and the start of
a new extended cycle. 

As the oppositely signed bands migrate toward the equator in the ``declining phase,'' each band has two global-scale connections, one intra- and one extra-hemispheric connection. Eventually those four bands mutually cancel each other around the equator and inhibit the production of sunspots, creating ``solar minimum'' conditions. The next phase occurs when, again, the equatorial bands cancel and spot growth starts again on the mid-latitude band. This perpetual interaction, or telecommunication, of the temporally offset 22-year long magnetic activity bands appear to drive the quasi-11-year cycle of sunspot production. 

Realizing that this cycling is taking place and that the onset of the high-latitude band is regular permits some level of predictability of the temporal landmarks of solar cycles. Information about strength would then appear to hinge on the strength in individual bands and how those bands interact inside a hemisphere and across the equator. In recent cycles, however the termination point has been spanning out in time, meaning that solar minimum conditions exist for longer as the period of mutual cancellation is extended. This also has the effect of reducing the length of the ascending phase. It is unclear what the conclusion of this phasing is, with systematically reducing cycle peak amplitudes, a scenario that is explored in a separate paper \citep[][]{2015FrASS...2....2M}.

Finally, in an extension of their work on BPs, \citet{McIntosh2017} combined BP analyses from the {\em Solar Dynamics Observatory} (\sdo) Atmospheric Imaging Array \citep[AIA;][]{2012SoPh..275...17L} and twin Exteme-UltraViolet Imagers \citep[EUVI;][]{2008SSRv..136...67H} of the {\em Solar-TERrestrial Relations Observatory} (\stereo) to build a picture of global-scale magnetism that spanned the entire solar atmosphere from 2011 to 2014. The features observed in longitude vs time plots in bands of constant latitude are consistent with the presence of magnetized Rossby waves on the same activity bands that were the likely source of strong quasi-annual outbursts of activity \citep[e.g.,][]{2015NatCo...6E6491M}. Further, in studying these global-scale phenomenon, they noted that the structures exhibiting enhanced activity persisted on very slow ($\sim$3~m/s) westward migrating solar longitudes that appeared to be stably separated in longitude for long periods of time, or that those waves had relatively stable longitudinal wavenumbers for the period studied \-- a fact that we utilize later in this Letter.

The analysis of \citet{2014ApJ...784L..32M} concluded with an assessment of the magnetic bands responsible for solar cycle 24 and what the patterns discovered could lead us to deduce about the appearance of their solar cycle 25 counterparts. In the following section we will recap that deduction before updating the diagnostic diagrams created some three and a half years later that motivate this Letter and the clear identification of the solar cycle 25 magnetic activity bands. Subsequent sections discuss what we could have stated about the hemispheric sunspot maxima of solar cycle 24, based on the forecast of \citet{2014ApJ...784L..32M}, before exploiting the slowly evolving active longitudes present in the current mid-latitudes to enhance our AIA diagnosis and indicate that our forecast is on-track and that solar minimum conditions are imminent.

\section{Recap: Forecasting Cycle 24 Evolution and Cycle 25 Onset}\label{recap}
\begin{figure}[!htb]
\plotone{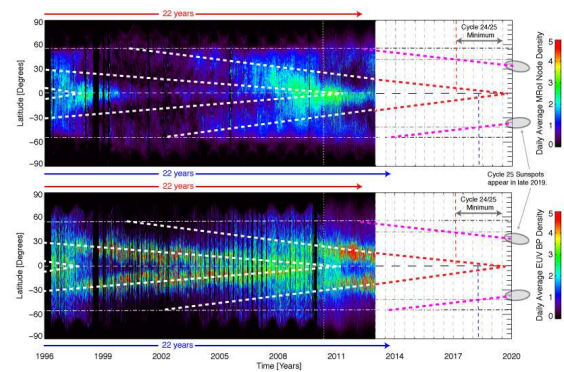}
\caption{The extrapolation of the inferred magnetic activity bands based on the analysis of \citet{2014ApJ...784L..32M} (Figs.~6 and 17). The white dashed lines are fitted to the latitudinal maxima of the daily hemispheric g-node (top) and BP (bottom) density functions that illustrate the complete migratory progress of solar cycle 24. The red-dashed lines are the linear continuation of the solar cycle 24 bands. The projected onset of the solar cycle 25 activity bands appear as pink dashed lines starting at high latitudes 22 years from the start of their cycle 23 counterparts. The faint vertical dashed gray lines in the panels are 6 months apart. \label{f1}}
\end{figure}

\cite{2014ApJ...792...12M} developed a ``forecast'' of solar activity in the near future using the measured progression of the magnetic activity bands of cycle 24, projected onset times of new activity bands at high latitude, and the average migratory speed from high-to-low latitudes as a guide. This forecast was produced to add some intellectual validity to the proposed data-motivated picture and indicated that the cycle 24 bands would terminate at the equator in late 2019. A further component in this forecast was the inference of when the cycle 25 bands would appear and the migratory speed \-- the pink dashed lines  in Fig.~\pref{f1}. The forecast was made under the assumption that the activity band migration continues at a constant speed and the speed assigned to the bands of cycle 25 was the average of those derived from data stretching back to 1859. Further, that the onset times of the new cycle bands appeared $\sim$ 22 years after the band of the same polarity in that hemisphere appeared \-- a time that defined the hemispheric maxima as we have noted above. All-in-all we projected that, as is usually the case, a few spot groups of cycle 25 might appear before the termination point, but that significant cycle growth would commence in earnest only after that time. 

\subsection{On the Occurrence of Hemispheric Sunspot Maxima}\label{hemispheres}
\begin{figure}[!htb]
\plotone{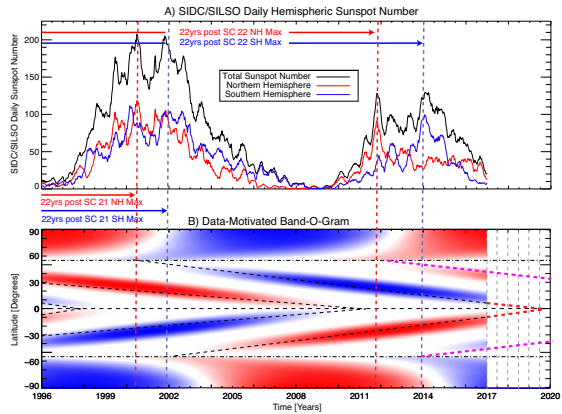}
\caption{Comparing the evolution of the daily hemispheric sunspot number (top) and a data-inspired representation of activity band polarity and migration (bottom). The top panel shows the variation of the daily sunspot number in the northern (red) and southern (blue) hemispheres while the total sunspot number is represented in black. The northern and southern hemispheric maxima are indicated as red and blue dashed vertical lines, respectively. \label{f2}}
\end{figure}

The forecast of \cite{2014ApJ...792...12M} missed a significant opportunity to explicitly declare when the hemispheric maxima of solar cycle 24 would occur based on our best estimates of when the bands of solar cycle 25 would appear at high latitudes. In Fig.~\pref{f2} we show the hemispheric maxima of solar cycles 23 and 24 in comparison with a ``band-o-gram'' or data-inspired schematic of activity band strength, position, polarity and interaction, see Fig.~8 of \citet{2014ApJ...784L..32M}. As discussed above, we discovered that the hemispheric maxima occur approximately 22 years after the band of the same polarity in that hemisphere and so, we could have simply declared that the hemispheric maxima would occur in late 2011 and that the southern maximum would occur in late 2013. Indeed, such a projection would have been borne out by the observations.

\section{Magnetic Cycle Diagnostic Update}\label{update}
\begin{figure}[!htb]
\plotone{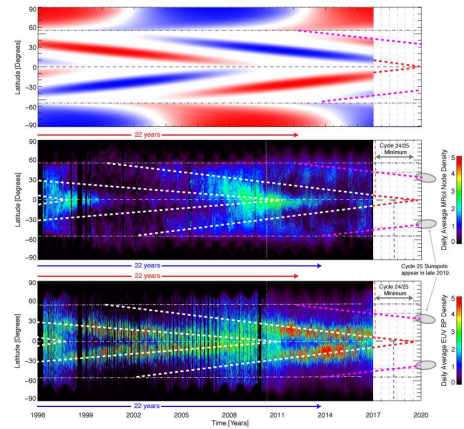}
\caption{As for Fig.~\pref{f1} with the g-node (center) and BP (bottom) density diagnostics updated to the end of 2016 and the updated band-o-gram (top). We see that the magnetic bands of solar cycle 25 not only appeared where and when anticipated, but that they appear to be following the mean migratory speeds inferred by \citet{2014ApJ...784L..32M}. \label{f3}}
\end{figure}

Figure~\pref{f3} brings our daily g-node, BP diagnostics, and band-o-gram picture up to date. The dashed lines shown have not been refitted to the g-node and BP distributions for ease of comparison with Fig.~\pref{f1}. As for cycle 23, we see that a prevalent trail of high g-node density is visible at high latitudes and marks the early phases of the extended cycle in both hemispheres, where the bands start on time {\em and} closely follow the pink dashed lines that represent their migratory behavior. Further, we see that their migration start is offset between north and south considerably as was projected. Therefore, it would appear that deductions of \cite{2014ApJ...792...12M} with respect to solar cycle 25 are on-track.

We note, as was also seen in the timeframe between 2006 and 2011, that there is notable growth in BP density as the four activity bands come into close conjunction on the approach to solar minimum. At present there is an enhancement in the BP density on the pink dashed line in the northern hemisphere (visible since the start of 2016) while the southern BP density enhancement has not yet become clearly distinguishable.

\section{Refining Daily Synoptic BP Charts}\label{synoptic}
The ability to clearly identify the bands of the upcoming cycle and characterize them in the future motivate this section. Their identification and characterization impact our ability to project what may happen, as we have seen above. 

\begin{figure}[!htb]
\plotone{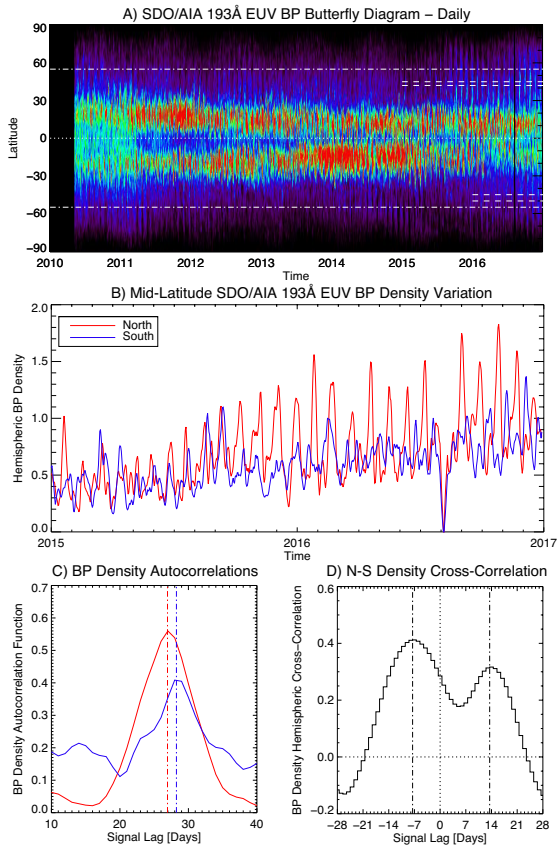}
\caption{From the latitude versus time plot of \sdo/AIA EUV BP density (panel A) where we identify mid-latitude regions, indicating the presence of solar cycle 25's magnetic activity bands with dashed horizontal lines. The extracted hemispheric timeseries are shown in panel B with the northern variation in red and southern in blue. Panel C shows the auto-correlation of each mid-latitude hemispheric BP density timeseries where the dot-dashed vertical lines indicate the times of peak recurrence in each signal at 27 and 28 days for the northern and southern hemispheres respectively. Panel D shows the cross-correlation between the hemispheric BP density profiles to illustrate that the disturbances present are not at the same solar longitude. \label{f4}}
\end{figure}

Close inspection of a higher resolution SDO/AIA daily BP density butterfly diagram (Fig.~\pref{f4}A) shows some interesting features that may indeed help us to refine the identification of the cycle 25 activity bands as they continue their migration and grown in strength. There is an apparent rotational-scale periodicity in the occurrence of BPs towards the end of the timeframe studied.
Those are best seen in the butterfly diagram at high latitudes in both hemispheres, between 40 and 45 degrees in the north and 45 and 50 degrees in the south. The difference in latitudes between the hemispheres stems from the $\sim$18 month temporal offset in starting the equatorward progression of activity bands from 55\degree{} between the hemispheres as seen in Fig.~\pref{f1} (and, in cartoon form, in Fig.~\pref{f2}). 

Again, the different sampled latitudes are a result of the temporal offset in hemispheric band migration (and hence hemispheric sunspot maximum) as mentioned above. Panel B of Fig.~\pref{f4} shows the relevant BP density timeseries averaged over that narrow range of latitudes in the northern (red) and southern (blue) hemispheres, and the quasi-periodic nature of both signals is clear. In order to assess the apparent periodicity in BP production and the apparent shift in time between the two hemispheres, we study their auto- and cross-correlation in panels C and D of Fig.~\pref{f4}. 

Panel C shows the auto-correlation for each hemisphere, again the north in red and south in blue. We see that the auto-correlation functions are relatively strong and relatively narrowly peaked at 27 and 28 days, respectively. This would imply that the underlying magnetic structures, the magnetic activity bands, have a small range of longitudes that are more active than others. In the context of the magnetized Rossby wave system interactions with solar activity that were recently identified \citep[][]{McIntosh2017}, this would relate to a ``wavenumber one'' disturbance that enhanced regions exist at one longitudinal location and that if that location migrates in longitude, its migration is very slow. Interestingly, the standard magnetic differential rotation periods would be 27.9 days at 43.5\degree{} latitude and 28.5 days at 47.5\degree{} latitude and so this system would appear to be at least one day faster than that in each hemisphere \-- possibly reinforcing the conclusion of \citet{2014ApJ...784L..32M} that BPs are tracers of magnetism rooted deeper in the Sun's convective interior.

Further, from the timeseries we see that the peaks in BP production do not necessarily happen at the same longitude in each hemisphere. Comparing the timeseries by using the cross-correlation of the timeseries between 2016 and 2017 we get the information presented in panel D, a double-peaked function at approximately 7 days behind and 14 days ahead. This would appear to indicate that, at these latitudes, the enhancing disturbances are separated by about 21 days of rotation or about 270\degree{} longitude. 

\begin{figure}[!htb]
\plotone{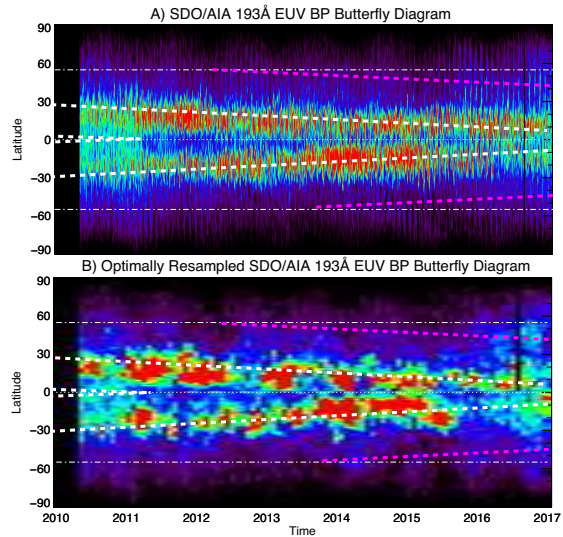}
\caption{Comparing the \sdo/AIA BP density butterfly diagram (top; see e.g., Figs.~\pref{f1} \&~\pref{f2}) with an optimized butterfly diagram (bottom) constructed using the properties derived from Fig.~\pref{f4}. In each panel we show the white and pink dashed lines that form the guidelines for the extended solar cycle, cycles 24 and 25 respectively. \label{f5}}
\end{figure}

Using this information we resample the daily AIA BP butterfly diagram for the appropriate interval in the northern and southern hemisphere to highlight the presence of the higher latitude bands. Sampling the northern hemispheric data every 27 and southern hemisphere every 28 days, with a north-south offset of 21 days, we recover panel B of Fig.~\pref{f5}. The panels of Fig.~\pref{f5} show, for reference, the dashed lines representing the bands of cycle 24 and 25. This resampling of the butterfly diagram would appear to strengthen our detection of cycle 25 bands in each hemisphere although we note that the southern hemispheric band is clearer, but also clearly reduced in strength. Further, this analysis would indicate that taking detailed observations of targeted longitudes could go a long way to monitoring cycle 25 and supporting the hypothesis of \cite{2014ApJ...792...12M} or refuting the same. Targeted observational plans are being constructed with the \iris{} and \hinode{} mission teams to monitor the progression of the cycle 25 bands and the final throes of the cycle 24 bands at the equator. Such investigations should be joined by the DKIST as it becomes operational around the same time as cycle 25 onset. 

\section{Conclusion}\label{theend}
We have presented observations that bring the diagnosis of \citet{2014ApJ...792...12M} to the current time. Those diagnostics indicate that the magnetic activity bands that will give rise to solar cycle 25 are visible and follow the evolutionary paths anticipated in the earlier analysis. The diagnostics follow these paths to a point where the earlier projection of solar cycle 25 onset, in the form of the first few spots in each hemisphere, in late 2019 or early 2020 would appear to be on track. The projected termination of the solar cycle 24 bands around the same time indicates that the time would mean an ascending phase of cycle 25 that is about 2 years long, depending again on the  22-year time between the onset of the cycle 24 and 26 bands in around 2022 at high latitudes. It remains to be seen why the polar circulation appears to be so robust, permitting the success of this forecast and, for solar cycle forecasts in general, \citep[e.g.,][]{2016SpWea..14...10P}. 

We anticipate that a short ascending phase would appear to favor a weaker cycle 25 (than 24; cycle
24's ascending phase was shorter than that of cycle 23) as there is more overlap time between the oppositely signed bands. Other authors, exploiting different means for forecasting solar cycle strength and landmarks \citep[e.g.,][]{2016ApJ...823L..22C, 2016JGRA..12110744H}, expect cycle 25 to be similar in magnitude to the cycle 24, which is still relatively weak in historical terms. The diagnostic technique of targeting prevalent longitudes for detailed observation may enhance our ability to monitor the onset and growth of cycle 25 as never before. 

\acknowledgments
The National Center for Atmospheric Research is sponsored by the National Science Foundation. The compilation of feature databases used was supported by NASA grant NNX08AU30G and the National Science Foundation. The imaging data used in this paper are freely available from the \sdo{} and \stereo{} mission archives and the Virtual Solar Observatory (VSO; http://virtualsolar.org)


\end{document}